\documentclass[a4paper,fleqn,usenatbib,useAMS]{mnras}
\usepackage{mathptmx}

\usepackage[T1]{fontenc}
\usepackage{ae,aecompl}
\usepackage{aecompl}

\usepackage{graphicx}	% Including figure files
\usepackage{amsmath}	% Advanced maths commands
\usepackage{amssymb}	% Extra maths symbols

\usepackage{times}
\usepackage{epstopdf}
\usepackage{graphicx}
\usepackage{txfonts}
\usepackage{natbib}
\usepackage{array}
\usepackage{verbatim} 
\title[Edge-on disc effect in B0712+472]{SHARP - IV. An apparent flux ratio anomaly resolved by the edge-on disc in B0712+472}

\author[Hsueh et al.]{J.-W. Hsueh$^{1}$\thanks{E-mail:
jwhsueh@ucdavis.edu} L.  Oldham$^{2}$\thanks{E-mail: loldham@ast.cam.ac.uk},  C. Spingola$^{3}$, S. Vegetti$^{4}$, C. D. Fassnacht$^{1}$, M. W. Auger$^{2}$,
\newauthor  L. V. E. Koopmans$^{3}$, J. P. McKean$^{3,5}$, and D. J. Lagattuta$^{6}$ \\
$^{1}$Physics Dept., University of California, Davis, 1 Shields Ave.
Davis, CA 95616, USA\\
$^{2}$Institute of Astronomy, University of Cambridge, Madingley Road, Cambridge, CB3 0HA, UK\\
$^{3}$Kapteyn Astronomical Institute, University of Groningen, P.O.Box 800, 9700AV, Groningen, the Netherlands\\
$^{4}$ Max Planck Institute for Astrophysics, Karl-Schwarzschild-Strasse 1, D-85740 Garching, Germany\\
$^{5}$Netherlands Institute for Radio Astronomy (ASTRON), P.O. Box 2, 7990 AA Dwingeloo, The Netherlands\\
$^{6}$ University of Lyon, CRAL, Observatoire de Lyon, 92 Rue Pasteur, 69007 Lyon, France
}
\begin{document}

%\date{Accepted 1988 December 15. Received 1988 December 14; in original form 1988 October 11}

\pagerange{\pageref{firstpage}--\pageref{lastpage}} \pubyear{2016}

\maketitle

\label{firstpage}

\begin{abstract}
Flux ratio anomalies in quasar lenses can be  attributed to dark matter substructure surrounding the lensing galaxy and thus used to constrain the substructure mass fraction.  Previous applications of this approach infer a  substructure abundance that is potentially in tension with the predictions of $\Lambda$CDM cosmology. However, the assumption that all flux ratio anomalies are due to substructure is a strong one, and alternative explanations have not been fully investigated. Here, we use new high-resolution near-IR Keck~II adaptive optics imaging for the lens system CLASS~B0712+472 to perform pixel-based lens modelling for this system and, in combination with new VLBA radio observations, show that the inclusion of the disc in the lens model can explain the flux ratio anomalies without the need for dark matter substructures. The projected disc mass comprises 16\% of the total lensing mass within the Einstein radius and the total disc mass is $1.79 \times 10^{10} ~ M_{\sun}$. The case of B0712+472 adds to the evidence that not all flux ratio anomalies are due to dark subhaloes, and highlights the importance of taking the effects of baryonic structures more fully into account in order to obtain an accurate measure of the substructure mass fraction.
\end{abstract}

\begin{keywords}
gravitational lensing -- quasars: individual CLASS B0712+472 -- galaxies: structure 
\end{keywords}

\section{Introduction}
Simulations based on $\Lambda$CDM cosmology predict that every galaxy halo should be populated by an abundance of low-mass substructure \citep{Klypin1999,Moore1999,Springel08}. However, such a large satellite population has not yet been identified in the Local Group \citep{S07,DES15,Kop15}. One explanation may be that haloes of such low masses are unable to form stars and are therefore dark \citep[e.g.,][]{Klypin1999,Bullock2000,P11}; if true, this prediction must be tested by probes that are sensitive to the mass, rather than the light. Strong gravitational lensing is therefore an excellent way to investigate the possibility of dark matter substructure in systems outside the Local Group.  In particular, when a quasar is multiply imaged by an intervening massive galaxy, any deviations of the flux ratios of the different images from the predictions of a smooth lensing mass model may be due to dark matter substructures that perturb the lens potential at the positions of the quasar images \citep{Mao1998,metcalf01}. Observations of these \emph{flux ratio anomalies} may therefore provide a key test of the $\Lambda$CDM cosmological model against alternative dark matter models, in which substructure is suppressed by a turnover in the primordial matter density power spectrum \citep[e.g.,][]{Vog2016,Sawala2016}. 

The first application of this method was carried out by \citet{Dalal2002}, in which seven
% Yes, D&K only looked at 7 lenses - B1555 was not part of their original sample.
radio-loud AGNs were used to constrain the projected substructure abundance to  0.6 - 7 \% around the critical curve of the host galaxies. Later studies of flux ratio anomalies have applied a strategy similar to that of \citet{Dalal2002}, modelling lensed quasar systems with a smooth mass model and attributing deviations of the observed flux densities from the predicted values to dark substructure \citep[e.g.,][although \citealt{Nierenberg17} rules out the previous substructure detection in HE0435-1223]
{Fassnacht99,Bradac02,Fadely2012,N14,Nierenberg17}.  
% Tried to add B2045 discovery paper to this cite but overleaf won't let me. ????
Radio-loud systems are especially suitable for this type of analysis because radio wavelengths are much less strongly affected than optical wavelengths by dust extinction and stellar microlensing \citep[although see also][]{KdB2000}. The inferred substructure mass fraction from flux ratio anomalous lenses in \citet{Dalal2002} is marginally consistent with both cold-dark-matter-only simulations \citep{Xu15} and the constraints derived by \citet{V14a} after applying the independent \emph{gravitational imaging technique} \citep{K05,V09} to a different sample of lenses. However, with the limited sample size and the uncertainties from intrinsic variation of quasar fluxes, the inferred substructure mass fraction in \citet{Dalal2002} does not provide a tight constraint for distinguishing different species of dark matter.

%The current inferred substructure mass fraction from flux ratio anomalous lenses is larger than, but marginally consistent with both cold-dark-matter-only simulations \citep{Xu15,Despali2016} and the constraints derived by \citet{V14a} applying the independent \emph{gravitational imaging technique} \citep{K05,V09} to a different sample of lenses.  

A detailed comparison with dark-matter-only simulations concluded that there was only a 1--4\% probability that dark-matter substructure has produced the observed strength of the observed flux ratio anomalies in the sample of lenses analysed by Dalal \& Kochanek and subsequent studies \citep{Xu15}, suggesting that other effects, such as the use of overly simple mass models \citep[see][]{evans03} for the primary lens that  incorrectly predict the expected image flux densities, may also play a significant role. 
The analysis of the EAGLE \citep{eagle} and Illustris \citep{illustirs} hydro-dynamical simulations by \citet{Despali2016} has shown that the expected number of low-mass substructures can be significantly affected by baryonic processes, and may be further reduced by 20 - 40\% (depending on the details of feedback processes) -- potentially making the discrepancy larger \citep[also see][]{Chua2016}. These results from the simulations show that the substructure population may not be able to generate the current observed flux ratio anomaly strengths in several lens systems and, thus, may imply alternative causes for the observed anomalies. 
%\citet{Wetzel2016} have also shown that high-resolution hydro-dynamical simulations can reproduce the observed number of low-mass luminous satellites in a single Milky Way-mass halo. 

Since the original \citet{Dalal2002} study, observational facilities have developed such that it is now possible to observe the anomalous systems in much more detail and, thus, to explore alternate explanations for their observed anomalies. One of these is the presence of a disc in the lensing galaxy where, especially if the disc is edge-on or close to it, the disc can provide the small-scale perturbations in the projected mass distribution that lead to flux-ratio anomalies. For example, \citet{Hsueh2016} used deep adaptive-optics-assisted (AO) observations with NIRC2/Keck to reveal the presence of a faint, previously undetected edge-on disc lying across the merging images of the gravitational lens system B1555+375.  In this system, a lens model that included the disc could reproduce the observed flux ratios without requiring additional substructure.
%and showed that the inclusion of the disc in the lens model removed the need to invoke the presence of dark substructure. 
Dark-matter N-body simulations have shown that interactions with a disc component in Milky Way-sized halos can lead to a depletion in the number of substructures \citep{Errani2017,DO2010,Yurin2015,Zhu2016}. Thus, disc lenses may have a low substructure mass fraction compared to elliptical galaxies and, therefore, an increased  probability that a feature such as an edge-on disc is causing an observed flux anomaly.
%The dark-matter-only N-body simulations have also shown the depletion in substructures due to the interaction with the disc in the Milky Way-sized halo \citep{Errani2017,DO2010,Yurin2015,Zhu2016}. These studies may imply a general phenomena that the presence of a disc results in a lower substructure mass fraction compare to the halo hosts a elliptical. The edge-on disc acting as a source of flux ratio anomaly is supported by the simulation results since the current edge-on disc lens systems like B0712+472 and B1555+375 have strong anomaly strength but theoretically have less substructure population. 
A similar situation
%, whereby flux-ratio anomalies may not be caused by substructure, 
can arise from complex baryonic structure in elliptical galaxies.  \citet{Gilman2016} used observations of nearby galaxies with detailed luminous mass distribution information to generate simulated lenses, and showed that structures in the stellar distribution can contribute up to $\approx 10 \%$ of the anomalies. 

In this work, we present new observations of the gravitational lens system CLASS B0712+472 \citep{jackson98} and show that this system represents another case where the observed flux ratio  anomaly can be explained by the presence of  an edge-on disc.\\
This paper is structured as follows. In Section 2, we present our new near-IR and radio datasets, which we use to introduce a revised lens model in Section 3. In Section 4, we discuss our results and their potential implications for the quantification of mass substructures using flux ratio anomalies. Throughout the paper, we assume a flat $\Lambda$CDM cosmology with $\Omega_m=0.27$ and $h=0.7$. \\
 
 %===========================================================
 
\section{Observations \& Data reduction}

The gravitational lens system B0712+472 has four lensed images with a maximum separation of 1.274~arcseconds in a fold configuration (see Fig.~\ref{fig:b0712radio}). It was discovered as part of the Cosmic Lens All-Sky Survey \citep[CLASS;][]{CLASS1,CLASS2}, as reported by \citet{jackson98}. Spectroscopy of the system revealed that the lens galaxy is at redshift $z_\ell = 0.41$, while the background object is a radio-loud AGN at $z_s = 1.34$ \citep{Fassnacht98}. The high-resolution imaging with the Hubble Space Telescope (HST) showed that the lensing galaxy has a highly inclined disc, whilst the host galaxy of the background AGN is lensed into extended arcs in the optical and near-infrared (NIR) imaging. The dataset that we use for the models presented in this paper consists of optical and NIR imaging taken with the W. M. Keck-II Telescope and HST, and high-resolution radio imaging taken with the Very Long Baseline Array (VLBA).  These data are discussed below and summarized in Table~\ref{tab:obs}.  We also use the flux ratios measured by \citet{K03} from a six-month-long MERLIN monitoring campaign. 

%--------------------------------------------------------

\subsection{Keck LGS adaptive optics and {\it HST} imaging}

\begin{figure*}
\centering
\includegraphics[width=\textwidth]{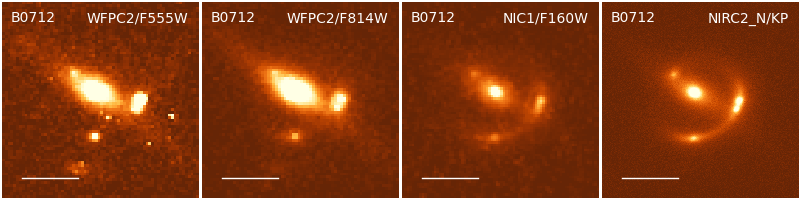}
   \caption{
   Multiband imaging of the B0712+472 system.  In each plot the white scale bar represents 1 arcsecond.
   {\bf (a)} Archival HST Wide-field/Planetary Camera 2 (WFPC2) image obtained in the F555W filter (roughly V-band).
{\bf (b)} Archival HST WFPC2 image in the F814W filter (roughly I-band). 
{\bf (c)}  Archival HST Near-infrared Camera and Multi-object Spectrograph (NICMOS) image obtained in the F160W filter (roughly H-band). 
 {\bf (d)} Keck AO image obtained in the $K^\prime$ band.
 }\label{fig:multiband}
\end{figure*}

B0712+472 was observed as part of the Strong Lensing at High Angular Resolution Program (SHARP; Fassnacht et al., in prep) using the NIRC2 camera on the W. M. Keck-II Telescope on the night of 2001 Dec 2 (PI: Fassnacht), using the laser guide star (LGS) adaptive optics (AO) system with corrections derived from the laser guide star and a $R=17.2$-mag tip-tilt star located 7.5~arcsec from the lens.  The narrow camera mode was used, giving a field of view of $\sim10$~arcsec on a side and a pixel scale of 10~mas. The imaging consisted of 19 dithered 300~s exposures that were obtained in the $K^{\prime}$ filter.  The data were reduced with the standard SHARP pipeline, which is a python-based package that is a refinement of the process described by \citet{Auger_EELS1}. Both stacked images (including all exposures) and individual-exposure images were created, in order to make sure that the point-spread function (PSF) was well-understood for the lens modelling (see Section 3). A cutout of the final reduced image is shown in Fig.~\ref{fig:multiband} and also in Fig.~\ref{fig:AOresidual}a.
%, where the contours from 5-GHz Multi-Element Radio Link Interferometer Network (MERLIN) imaging by \citet{jackson98} and 1.65-GHz VLBA imaging (see below) are overlaid. 

B0712+472 was also observed with HST using the F555W and F814W filters (WFPC2; PI: Impey, GO-9138) and the F160W filter (NIC1/NICMOS; PI: Jackson; GO-7255). These data were reduced using the standard {\sc multidrizzle} pipeline, producing final drizzled images with a pixel size of 50~mas. The reduced images are also shown in Fig.~\ref{fig:multiband}. The lens and the AGN images are detected in the \textit{HST} optical bands, but the lensed host galaxy is only apparent in the Keck II $K^\prime$ band imaging. We therefore do not use the HST data in our lens modelling, but do show a multiband image 
%combining the F555W, F814W, F160W and $K^\prime$ band filters 
in Fig.~\ref{fig:multiband}.

\begin{table}
\centering
\caption{Summary of the B0712+472 observations.}
\begin{tabular}{llccc}
\hline
Telescope		& Camera			&  Band 		& Date				&$t_{exp}$ (s) \\
\hline
VLBA			&					& 1.65 GHz	& 	2016 Feb 13	& 27360\\
\textit{HST}	& WFPC2    		& F555W		&	2001 Oct 27	& 13300\\
\textit{HST}	& WFPC2    		& F814W		&	2002 Oct 07 	& 8000\\
\textit{HST}	& NICMOS/NIC1	& F160W		&	1997 Aug 24	& 5246\\
Keck-II			& NIRC2 AO		& $K^\prime$	& 	2010 Dec 02	& 5700\\
\hline
\end{tabular}
\label{tab:obs}
\end{table}

\subsection{Very Long Baseline Array imaging}

The B0712+472 system was observed with the Very Long Baseline Array (VLBA) on 2016 Feb 13 (BS251 project, PI: Spingola). The observations were carried out in full continuum mode at a central frequency of 1.65 GHz (L band) in full polarization, with a recording rate of 2 Gbps and 4 second integration time.
The total bandwidth is 256 MHz divided up into two IFs  with 256 channels of 500 kHz width in each polarization, and a total observing time of 12 hours. The observations were performed in phase referencing mode, with a regular switching between the target and the phase calibrator (J0720+4737, with angular separation of 1.38 degrees) 
of 4 minutes, resulting in a net observing time on B0712+472 of $\sim$ 7.6 hours.

\begin{figure*}
 \centering
	\includegraphics[scale = 0.4]{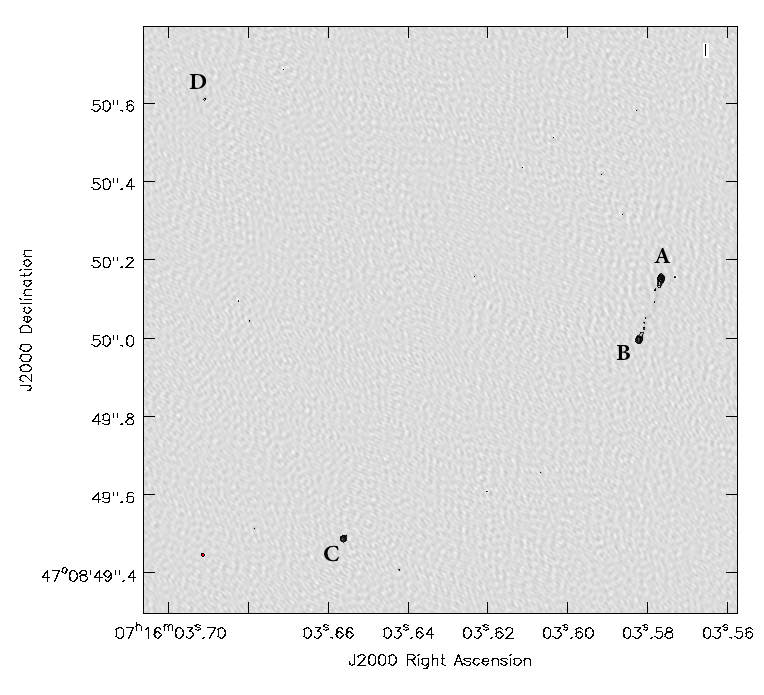}
    \includegraphics[scale = 0.42]{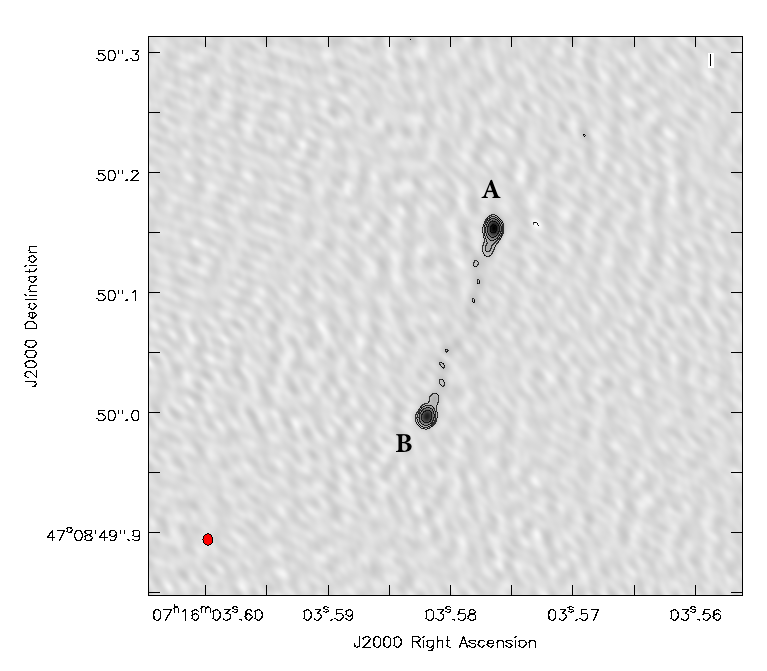}
   \caption{ \textsl{Left:} VLBA 1.65 GHz image of B0721+472. The rms is 31 $\mu$Jy beam$^{-1}$ and the peak is 6.3 mJy beam$^{-1}$. The restored elliptical Gaussian beam has a FWHM of 9.5 $\times$ 8.2 mas$^2$ in p.a. $5.3^\circ$ north to east and it is plotted in red in the bottom left corner of each image. The first contour intensity is the off-source noise level; contour levels increase by a factor of 2. \textsl{Right:} Zoom on the  images showing flux ratio anomaly, A and B. Note that there is a 1 sigma detection of a faint distorted arc between the images.}\label{fig:b0712radio}
\end{figure*}

We carried out a full calibration by using the Astronomical Image Processing System ({\sc aips}) software package. The data were reduced using the {\sc vlbautils} procedures following the standard calibration method as follows.
We first performed phase calibration by correcting for the Earth orientation parameters, the delay from ionosphere and parallactic angle. Then we applied an \textsl{a priori} amplitude calibration, which uses measured system temperatures and gain curves. We removed the residual instrumental phase and delay offset by using the information provided by pulse-cal tones table. The residual fringe rates and delays were determined by applying the fringe fitting procedure using the {\sc aips} task {\sc fring}. Finally, we used the primary calibrator DA193 for the bandpass calibration. No frequency or time averaging was undertaken during the calibration. This provided an effective field of view of $\sim$ 15 arcsec. 
Imaging was carried out with the {\sc clean} algorithm in {\sc aips} using natural weighting of the visibilities to improve the sensitivity to the extended emission from the jets, and restored using an elliptical Gaussian beam of size $9.5 \times 8.2$~mas$^2$ at a position angle of $5.3^\circ$ east of north. The final map (Fig.~\ref{fig:b0712radio}) has an rms of 31~$\mu$Jy beam$^{-1}$.

The total flux density of B0712+472 was determined in the image plane by placing an aperture over the area that contains the four components using the {\sc tvstat} task, and was found to be $21.9\pm1.8$~mJy. 
The gain corrections found during the calibration were of the order $\sim 8$ per cent, which
was assumed as a conservative estimate of the error on the absolute flux density.
The flux densities of the individual images were estimated by a Gaussian fit on the image plane using the {\sc jmfit} task.  The uncertainties on the flux density ratios are calculated as the product of the flux ratio with the square root of the sum of the squares of the fractional uncertainties on each of the flux densities. In Table \ref{tab:radio} we list the flux density ratios and their uncertainties.  

We find that the fold images A and B show extended core-jet structure in both images (Fig.~ \ref{fig:b0712radio}). There is a hint of a perturbed arc between A and B of $\sim165$~mas in length. Image C is still compact on mas-scales, whereas image D is only marginally detected ($\sim 1~\sigma$). 

\begin{table*}
\centering
\caption{The multi-band observed flux density ratios of B0712+472 lensed images. The MERLIN 15 GHz data are from \citet{jackson98}, the 5 GHz data are from the MERLIN Key Project \citep{K03}, and the VLBA 1.65 GHz and Keck II AO K'-band data are from this work. Note that the VLBA 1.65 GHz and MERLIN 15 GHz data are one-time measurements while the MERLIN 5 GHz data are the average flux ratios with rms scatter from a six-month-long monitoring programme.}
\begin{tabular}{ccccc}
\hline
Flux Ratio & VLBA 1.65 GHz & MERLIN 5 GHz & MERLIN 15 GHz & Keck AO $K^\prime$\\
\hline
B/A & $0.663 \pm 0.005$ &$0.843 \pm 0.061$ & $0.744 \pm 0.075$ & $0.51 \pm 0.02$ \\
C/A & $0.327 \pm 0.009$& $0.418 \pm 0.037$ & $0.378 \pm 0.036$ & $0.27 \pm 0.02$\\
D/A &$0.068 \pm 0.045$ &$0.082 \pm 0.035$ & $0.071 \pm 0.007$  & $0.08 \pm 0.02$\\
\hline
\label{tab:radio}
\end{tabular}
\end{table*}

%===========================================================
 
\section{Lens modelling} \label{sec:modelling}

B0712+472 has several notable features in the high-resolution AO and {\it HST} imaging. The lens galaxy has a highly inclined disc that lies close to the lensed images B and D (Fig.~\ref{fig:AOresidual}a). In \citet{Hsueh2016}, a similar configuration was revealed in B1555+375. In that system, a simple model that incorporated the newly-discovered edge-on disc could fully produce the observed flux ratio anomaly, without requiring the presence of substructures. Here, we investigate whether the same is true for B0712+472. 

\subsection{A simple mass model}

In \citet{Dalal2002} and \citet{Xu15}, the flux ratios were shown to be anomalous in the context of a simple mass model including  a singular isothermal ellipsoid (SIE) plus external shear. Before investigating more complex models, we here revisit the SIE+shear model in the context of our updated quasar image positions and monitoring flux-density ratios. We use the radio image positions (quasar image positions which are precise to 3 mas from our VLBA observations) and flux ratios from the six-month-long MERLIN monitoring of this lens by \citet{K03} as input constraints in {\sc gravlens} \citep{Kee01}. We do not include the flux ratios from VLBA 1.65 GHz and MERLIN 15 GHz in the lens modelling since a one-time flux measurement can be offset by up to 20 per cent from the average flux \citep[see fig. 1 in][]{K03}. The Einstein radius of the SIE+shear model is 0.64 arcseconds. We find that this model can reproduce the image positions in radio, but not the flux ratios, within 3-sigma measurement uncertainties. The model-predicted flux ratios are included in Table 4; the discrepancy between these predictions and the data then motivates our introduction of an explicit disc mass component in the following subsection.

\subsection{A mass model with a disc}
Motivated by the Keck AO and HST imaging, in which the disc is evident, we improve on the simple SIE+shear model of the previous section by explicitly adding mass in the form of an exponential disc. Thus our total mass model now consists of an SIE (which can be considered as accounting for the mass of both the dark halo and the luminous bulge), an exponential disc, and external shear. 

The second major improvement of our mass model in this section is that we now combine the position and flux constraints from the radio observations -- which in the past have been the \emph{sole} constraints on the mass model -- with our new $K^\prime$ imaging data, which allows us to perform pixel-based modelling in order to extract information from the full surface brightness distribution of the Einstein arcs. We are therefore able to obtain much more precise constraints on the lensing mass than has previously been possible for this system.

We combine the radio and optical datasets as follows. For the $K^\prime$ imaging, we simultaneously model the lensing mass and the light from the AGN host galaxy, AGN, and the lens galaxy, following the methods of \citet{Oldham2016}. We use a single S\'ersic profile for the host galaxy and decompose the lens galaxy light into two S\'ersic profiles to account for the bulge and disc light; because the optical flux densities of the AGN images may be affected by extinction and stellar microlensing, we model these as pseudo-foreground PSFs (i.e., with arbitrary fluxes, as opposed to determining them explicitly from the lens model), but with inferred \textit{positions} that are constrained by the lens model. As the AO PSF is uncertain and varies between individual exposures, we exploit the eight individual exposures in which an unsaturated star is visible and use this as the PSF in each case; our model therefore also includes further additional parameters describing the pixel-level offsets between the eight different frames. After we construct the initial lens model from the optical datasets, we then \emph{simultaneously} include the constraints from the radio data, using {\sc gravlens} to find the quasar image positions and flux ratios in the radio, and the quasar image positions in the $K^\prime$ band, for a given model, which we compare with the data in a chi-squared sense. Finally, we account for a registration offset between the radio and AO datasets.

\begin{figure*}
\centering
\includegraphics[width=\textwidth]{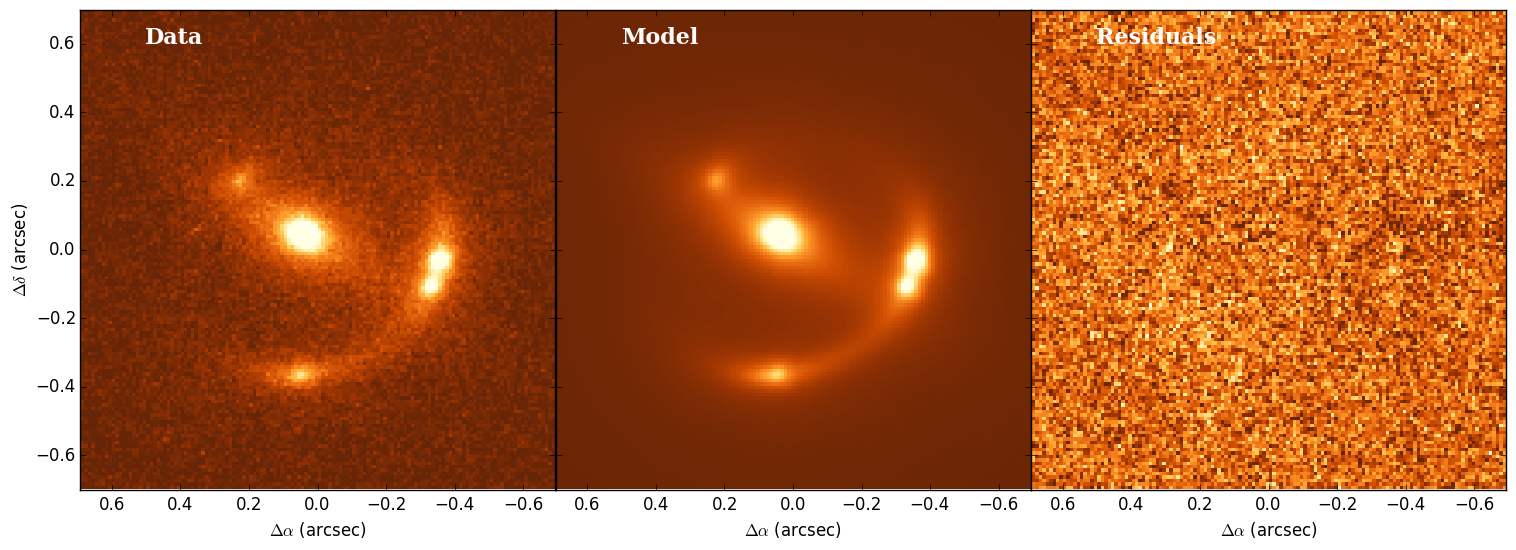}
   \caption{
   {\bf (a)} $K^\prime$ band image of the B0712+472 system obtained with the Keck-II adaptive optics system.  This is one of the eight exposures that was used in the lens modelling.
{\bf (b)} Best-fit lens model reconstructed image. 
{\bf (c)}  Residual image (data $-$ model).  
 }\label{fig:AOresidual}
\end{figure*}

Fig.~\ref{fig:AOresidual} shows the data, model and signal-to-noise residuals for one of the eight $K^\prime$ frames used for the modelling. We are generally able to recover the radio positions and flux ratios within $1 \sigma$ and describe the eight $K^\prime$ frames virtually down to the noise.
It is notable that although the optical pixels dominate the total number of constraints, the uncertainties on the flux ratios and positions are sufficiently small that the change in their contribution to the log likelihood is significant. Effectively, the optical pixels guide the model to zeroth order, and the radio data drives it to better discriminate between models within that space.

\begin{figure}
\centering
	\includegraphics[scale=0.5]{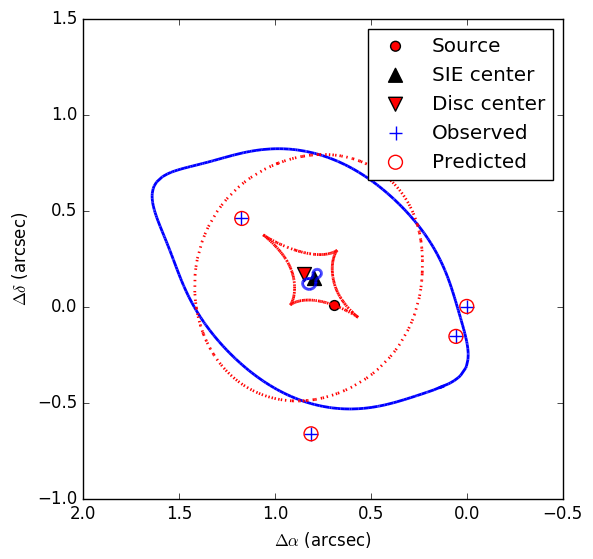}
   \caption{Observed radio positions: blue
pluses; Model-predicted image positions: red open circles; Lens-plane critical curves: blue solid curves; Source-plane caustics: red dotted curves. The source position is at (0.6953, 0.0152), which is marked by the red filled circle. The offset between extended source centroid positions the point source position is within the FWHM of AO PSF. The centroid positions of the SIE and exponential disc components are marked by the
black and red filled triangles respectively. 
   }\label{fig:lens}
 \end{figure}

\begin{table}
\centering
\caption{Best-fit model parameters and uncertainties of the two-component SIE + ExpDisc + external shear lens model. The positions are measured in arcseconds, with respect to image A in VLBA measurement. The position angles are measured in degrees east of north. The Einstein radius $b$ and disc scale radius ($R_s$) are measured in arcseconds.}
\begin{tabular}{ccc}
\hline 
Parameter    & SIE Component & ExpDisc Component  \\
%   & (SIE) & (ExpDisc)\\
\hline
$\Delta$RA	& $0.785 \pm 0.005$		& $0.896 \pm 0.005$ \\
$\Delta$Dec	& $0.142\pm 0.004 $		& $0.200 \pm 0.004$  \\
$b$ 			& $0.609 \pm 0.007$  		& ...  \\
$\kappa_0$ & ... &$0.294 ^{+ 0.040}_{- 0.035}$\\
$q$	  		& $0.84 \pm 0.01$			& $0.23 \pm 0.01$  \\
$\theta$ 		& $71.2 ^{+ 2.8}_{- 3.0}$			& $59.7 \pm {0.3}$ 	 \\
$R_s$			& ...  						& $0.389 ^{+ 0.022}_{- 0.020}$	 \\
$\gamma$ & $0.096 ^{+0.005}_{-0.004}$ & ... \\
$\theta_{\gamma}$ & $34.4^{+1.6}_{-1.2}$ & ...\\
\hline
\label{tab:model}
\end{tabular}
\end{table}

\begin{table*}
\centering
\caption{A summary of the observational constraints and the model-predicted results for a lensed point source. The lensed image positions from VLBA (in arcseconds) and average flux ratios \citep{K03} are with respect to image A. Uncertainties in the position measurements are $0.003$~arcseconds.}
\begin{tabular}{cccccccc}
\hline
Image	&\multicolumn{3}{c}{Observed} 	 	& \multicolumn{4}{c}{Model-Predicted}\\
		&\multicolumn{2}{c}{VLBA 1.65 GHz}	&{MERLIN 5 GHz} & \multicolumn{3}{c} {SIE+ExpDisc+Shear} & SIE+shear\\
		&East &North & Flux Ratio &East 	&North & Flux Ratio &Flux Ratio\\ 
\hline
A  &$0$    		&$0$			&  $1$ 				&$+0.001$  &$+0.000$	& $1$ 		& $1$\\  
B  &$+0.056$	&$-0.156$	& $0.843 \pm 0.061$ 	&$+0.057$ &$-0.156$		& $0.830$ 	& $1.031$ \\  
C  &$+0.812$ 	&$-0.663$	& $0.418\pm 0.037$	&$+0.812$ &$-0.663$		& $0.364$ 	& $0.303$\\  
D  &$+1.174$	&$+0.459$	& $0.082 \pm 0.035$ 	&$+1.172$ &$+0.459$		& $0.077$ 	& $0.058$\\  
\hline
\end{tabular}
\label{tab:results}
\end{table*}

\begin{table}
\centering
\caption{Light profile fitting parameters to the lens galaxy of B0712+472 system in $K^\prime$ band. The light profile consists of two S\'ersic profiles, each described by a centroid relative to image A $(\Delta RA, \Delta Dec)$, an effective radius $R_e$, S\'ersic index $n$, axis ratio $q$, and position angle $\theta$. The effective radii are measured along the intermediate axis; the intermediate-axis half-mass radius of the massive disc (Table 3) is $0.31\pm0.02$ arcsec.}
%The effective radius is measured along the intermediate axis; the corresponding major-axis \textit{scale} radius (i.e., the radius at which the surface brightness has dropped by a factor of $e$ from the peak, equivalent to the scaling used for the mass) is 0.43 arcsec.}
\begin{tabular}{ccc}
\hline
Parameter & Disc Component & Bulge Component\\
\hline
$\Delta RA$ & $0.801_{-0.007}^{+0.011}$  & $0.806 \pm 0.001$  \\
$\Delta Dec$ & $0.147_{-0.006}^{+0.007}$ & $0.151_{-0.002}^{+0.001}$ \\
$R_e$ & $0.288_{-0.006}^{+0.013}$ & $0.1273_{-0.014}^{+0.017}$ \\
$n$ & $0.883^{+0.070}_{-0.147}$ & $3.958^{+0.197}_{-0.331}$\\
$q$ & $0.232_{-0.013}^{+0.027}$ & $0.687_{-0.046}^{+0.002}$\\
$\theta$ &$59.7_{-0.5}^{+0.9}$ & $73.9_{-3.7}^{+0.8}$\\
\hline
\end{tabular}
\label{tab:light}
\end{table}

\subsection{Results}
Our model-predicted results for the AGN point source are shown in Table \ref{tab:results} with comparison to both the observed data and the single SIE+shear model of Section 3.1. The best-fit model parameters and 68 per cent confidence level (CL) errors are listed in Table~\ref{tab:model}, the light parameters in Table 5, and the lens model is illustrated in Fig.~\ref{fig:lens}. We find that adding an exponential disc to a smooth SIE model successfully reproduces the flux ratios in this system. The positions are all in agreement, and the flux ratios are consistent at the $2\sigma$ level, with excellent agreement now obtained for the two merging images A and B. The AGN source position is at $(0.6924, 0.0104)$. We find a negligible offset between the host galaxy and AGN position in the source plane, indicating that the two are cospatial. 

The model requires that the disc component makes up about 16 per cent of the total projected mass within the Einstein radius, which is $M_{\rm E} = 1.37_{-0.16}^{+0.19} \times 10^{10} ~ M_{\sun}$ (or $M_{\rm E} = 9.58_{-1.14}^{+1.30} \times 10^9 ~ M_{\sun} ~ h^{-1}$ ). The total rest-frame $K^{\prime}$ luminosities of the disc and bulge are $\log (L_{d}/L_{\odot}) = 10.71_{-0.05}^{+0.06}$ and $\log (L_{b}/L_{\odot}) = 10.84_{-0.02}^{+0.02}$, giving a bulge-to-disc light ratio $L_{b}/L_{d} = 1.34_{-0.12}^{+0.14}$ and a disc mass-to-light ratio $M/L = 0.35_{-0.06}^{+0.07}$. The latter is consistent with a 2 Gyr stellar population forming with a Chabrier stellar initial mass function, according to the stellar population models of \citet{Bruzual2003} -- though we note that the models used for the disc mass and light are not identical (the former is an exponential disc, whereas the latter is a S\'ersic profile). Nevertheless, the disc mass is well-aligned with the light, and the mass and light scale radii are comparable. Finally, we note an offset of 0.12 arcseconds (660 pc) between the disc and SIE centroid positions, and a strong external shear $\gamma = 0.096_{-0.004}^{+0.005}$, which we discuss further in \S\ref{ssec:env}.

%but note here that the inferred offset may contribute to this strong shear (is this true??).

%===========================================================
 
\section{Discussion \& Conclusions}
In this paper we have presented new high-resolution imaging of the B0712+472 gravitational lens system in the near-IR with Keck II adaptive optics and at radio wavelengths with the VLBA.  We have also presented a new lens model which includes a disc component.  These new data, and the new lens model based upon them, have led to a number of new findings.

\subsection{Flux-ratio anomalies can be also caused by discs}

Based on the near infrared imaging data presented in this paper, which shows the highly-inclined disc of the lensing galaxy at a high signal-to-noise ratio, we have developed a mass model that includes a disc component in addition to the traditional SIE that describes the overall halo. Such higher-order modifications to the smooth mass model have been suggested by previous studies by e.g., \citet{evans03,congdon05}.  Our disc + SIE model is able to reproduce the observed radio-wavelength flux ratios in the B0712+472 system without the need for additional dark-matter substructure.  Of course our model does not rule out a possible substructure contribution to the image flux densities, but it  demonstrates that a straightforward mass distribution that is motivated by the observed lensing galaxy morphology can explain the flux ratio anomaly in this system.  
Our work on the B0712+472 system is now the second case in which a disc component in the lensing galaxy can explain an observed significant flux-ratio anomaly in the fold images of a radio-loud lens system.  Our previous work on the B1555+375 system showed the same effect, where the addition of an edge-on disc, once again motivated by the AO imaging data, could reproduce the observed flux ratio anomaly without the need for an additional substructure \citep{Hsueh2016}.

The B0712+472 system was one of the seven radio-loud lenses on which the \citet{Dalal2002} substructure analysis was based, and B1555+375 has been used in many subsequent substructure analyses \citep{KD04,Keeton2005,Dobler2006,Xu15}. 
The current sample of systems in which the flux ratios  can be reasonably expected to be free of microlensing or extinction is small ($\sim$10 systems), and three of these systems (B0712+472, B1555+375, and B1933+503) show obvious discs.  In two of the three, as shown by our previous study \citep{Hsueh2016} and  the results in this paper, the observed discs can explain the observed anomalies. 
%The current sample of systems for which flux ratios can be reasonably expected to be free of microlensing or extinction is small ($\sim$10 systems), of which we have now seen discs that can explain the observed anomalies in two. 
In future studies, where larger lens samples will be available, it will be possible to make a a pre-selection to avoid disc lenses. 
However, the analysis from \citet{Gilman2016} has shown that even in elliptical lenses, baryonic structures can still contribute $10 - 15 \%$ of anomalies.  
Thus, our results suggest that approaches that treating the lens-galaxy mass distributions solely with standard SIE models may overestimate the incidence of substructure. 

%The dark-matter-only N-body simulations have shown the depletion in substructures due to the interaction with the disc in the Milky Way-sized halo \citep{Errani2017,DO2010,Yurin2015,Zhu2016}. These studies may imply a general phenomena that the presence of a disc results in a lower substructure mass fraction compare to the halo hosts a elliptical. 

\subsection{The importance of high-resolution imaging}

The models in this paper were based on constraints that would not have been available without high-resolution imaging.  The first aspect of this is the  characterization of the disc component in the lensing galaxy, without which we would have not been able to investigate the influence of the baryonic component on the lensed AGN fluxes with pixel-based lens modelling.  The optical spectrum of the lensing galaxy in the B0712+472 system shows the features typical of elliptical galaxies, with no obvious signs of emission lines that might indicate star formation associated with a disc component \citep{Fassnacht98}.  Based on the spectrum alone, it would have been quite natural to assume that the lens was an early-type galaxy and, thus, to model it only with the standard SIE or related mass distribution.  Instead, we see a clear disc component, which our surface-brightness fitting to the $K^\prime$-band AO data reveals to have a S\'{e}rsic index of 0.883.  Similar behaviour is seen in the ``EELs'' sources, which have been spectroscopically selected such that both the lens and the background object have early-type galaxy spectra.  Several of the sources were shown in HST and SHARP Keck AO imaging to have a prominent disc component associated with either the lens or the background source \citep{Oldham2016}.

%Additionally, the high-resolution AO imaging was critical for the modeling of the B0712+472 system because in those data the arcs due to the lensed host galaxy of the background AGN provided large numbers of constraints for the models. Thus, for the B0712+472 system we were able to use the pixelated modelling technique \citep{K05,V09} \com{no, these weren't used here...} in a way that was not possible in our previous analysis of the B1555+375 system, which the arcs are suffered with the severe extinction from the edge-on disc.

The high angular resolution radio data also plays a key role.  
The astrometric precision on milliarcsecond-scale provided by the new VLBA imaging presented here puts a stringent constraint on the image positions used for the models, while the MERLIN monitoring campaign \citep{K03} provides the flux density ratios that the models need to match.  
\citet{jackson98} has already noticed the failure in fitting to the radio image positions with an SIE-only lens model. Although the SIE+shear model reproduces the astrometry in new VLBA imaging, it favors a small axis ratio ($\approx 0.5$) and a larger external shear, which still leaves notable residuals in flux ratios.  
Although a massive substructure can also perturb the image positions and cause astrometric anomalies, we find no sign of luminous satellites in the high-resolution AO imaging. The failure of the traditional SIE model to predict properly the observed positions argues, once again, that a more sophisticated model is required for this system.

%The new VLBA imaging presented here provides extremely high-precision image position constraints for the models, while the MERLIN monitoring campaign \citep{koopmans00} provides the flux ratios that the models need to match. [Bridge between A and B in VLBA imaging]

%[Bridge between A and B in VLBA imaging]
%Moreover, from VLBA imaging, we have found evidence for an extended gravitational arc between the images A and B at 1$\sigma$ level. The arc shows a hint of a perturbed nature, which could be a signature of a perturbation due to the substructure or the mass distribution of the lens galaxy (e.g. an edge-on disc) \citep{Metcalf2002,metcalf01}. However, the signal-to-noise ratio of the data is currently insufficient to confirm whether the arc is distorted by substructure, or whether the brightness distribution of the arc is due to the intrinsic structure of the background source.

%\newpage

\subsection{The observed flux ratios are robust}

When using flux ratios to investigate substructure, it is important to ensure that the observed systems are as free as possible from the effects of microlensing and absorption or other propagation effects.  The first condition requires that the angular sizes of the lensed emitting regions are larger than the roughly micro arcsecond scale of the Einstein ring radii of individual stars in the lensing galaxy.  This is typically achieved in systems where the lensed AGN is radio-loud \citep{K03,Rumbaugh2015,Pillips2000,B04,Katz97,jackson98,Patnaik99,Patnaik92,Marlow99,Fassnacht02,Cohn2001,Biggs2000,mckean07}, is detected in mid-IR wavelengths where the emission is coming from the dust torus surrounding the AGN \citep[][although the dust torus can also contribute to microlensing, see \citealt{Stalevski2012}]{Chiba2005,ML2013,VA2016}, or where emission from the narrow-line emission region can be isolated through integral field spectroscopy \citep{Nierenberg17,N14,Metcalf2004}. Our $K^\prime$-band flux ratios are shown in Table \ref{tab:radio}. These ratios do not account for the time delays between the images, but the greater than 50 per cent inconsistency with the average flux ratios from the 5~GHz data indicates that microlensing or other propagation effects are likely influencing these flux ratios.

The second condition can also usually be met by observing at radio or mid-IR wavelengths, where extinction due to dust is not a significant effect. However, there can be propagation effects such as interstellar scattering and free-free absorption \citep{Mittal07,winn04}, which appear to contribute at least in part to the strong flux ratio anomaly in the B0128+437 system \citep{B04}; or in some rare cases, micro-lensing from stars \citep{koopmans00}.  Because these effects have a well known power-law dependence on wavelength, it is possible to test for their presence by obtaining multi-frequency radio observations. For B0712+472, the multi-band radio flux ratios are shown in Table \ref{tab:radio}. The VLBA 1.65 GHz and MERLIN 15 GHz data are from one-time observations and therefore the flux ratios will have larger uncertainties than those listed in Table 2 due to the intrinsic variation of the AGN \citep[VLA monitoring at 8.5~GHz finds $\sim$5 per cent variability;][]{Rumbaugh2015}, but nevertheless the observations show no significant sign of power-law frequency dependency as propagation effects would cause. Thus, the data suggest that the radio wavelength flux ratios are not affected by propagation effects and, therefore, that the flux ratios from long-term monitoring used in our lens modelling are robust. 

%Our results demonstrate the power of high-resolution imaging which can greatly improve the performance of the lens model. B0712+472 was originally classified as an early-type elliptical with spectra \cite{Fassnacht98} but then revealed a clear edge-on disc in the AO imaging. The light profile fitting result in $K^{\prime}$ band has the best-fit S\'ersic index to be \com{**}. For the intention to avoid spiral lenses in the flux ratio anomaly analysis, the case of B0712+472 implies that a high-resolution optical/NIR imaging is still needed for gathering morphology information. 
%The light profile fitting result in $K^{\prime}$ band has the best-fit S\'ersic index to be \com{**}.

\subsection{The role of B0712+472 environment} \label{ssec:env}

\begin{figure}
%\centering
	\includegraphics[scale=0.55]{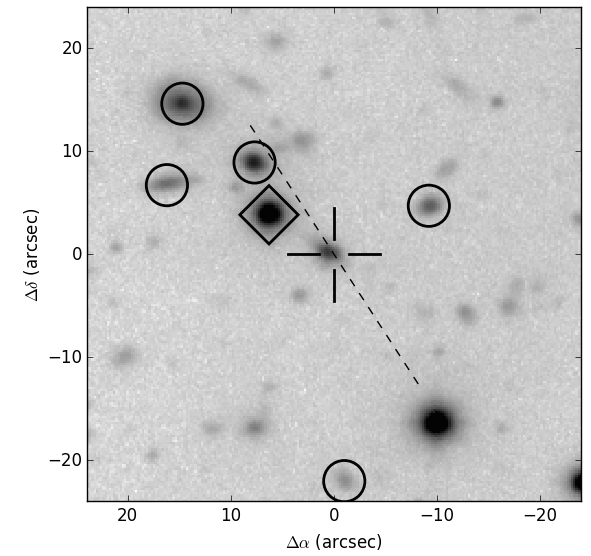}
   \caption{
   Local environment of the B0712+472 lens system, which is marked by the open plus sign.  The dashed line represents the position angle of the shear term in the lens model.  Circled galaxies are all part of a foreground structure at $z = 0.29$ \citep{Fassnacht02}, while the diamond marks the tip-tilt star used in the Keck AO observations of this system.  The figure is based on the Keck LRIS imaging described in \citet{Fassnacht02}.
   }\label{fig:environment}
 \end{figure}

In \S\ref{sec:modelling}, we found that our lens models required a large value for the shear ($\gamma \approx 0.1$) compared to the SLACS lenses which only need very little shear \citep{SLACS2006}. As part of our exploration of various lens models, we also constructed an alternative model with no shear, but found that a $\sim 2$ kpc offset between the SIE and disc mass centroids was required in this case.  In other words, this second model is compensating for the lack of shear by splitting the mass components in a manner that is unlikely to form a static disc.  Previous explorations of the environment of the B0712+472 system have found a foreground group of galaxies at $z=0.2909$ \citep{Fassnacht02}.  In particular, this group shows a tight concentration just to the northeast of the lens system (Figure~\ref{fig:environment}), in a direction that is consistent to the shear position angle in the lens model. To identify the most significant perturbers in the field quantitatively, we calculate their \textit{flexion shift} \citep{Mccully2014} and find that the largest source of the external shear are the three galaxies in the foreground group at $z=0.2909$ that are seen to the northeast of the lens system in Figure~\ref{fig:environment}.  
%Only one of these three galaxies exceeds the $10^{-4}$ value of the flexion shift that is often used to designate which galaxies need to be included in lens modeling
%\citep[e.g.][]{Sluse2016}, while the other two have values close to this cutoff, with flexion shift values of ?? and ??.  
This direction is also roughly aligned with the observed disc position angle.  Overall, it is likely that a combination of internal and external shear contribute to the large shear value seen in the modelling.

\subsection{Future work}

Constraints on the abundance of substructure will soon achieve a new stage of accuracy due to the thousands quadruply-imaged quasar lenses that are expected to be discovered by future surveys, such as the Dark Energy Survey \citep[DES;][]{DES2016}, \textit{Euclid} \citep{Cimatti2012} and the Large Synoptic Survey Telescope \citep[LSST;][]{LSST}. However, our results demonstrate that
in order to fully take advantage of these new large samples, high-resolution and high signal-to-noise ratio imaging will be needed to provide correct morphological information and exclude systems with prominent disc components. At the same time, as some massive elliptical lenses also show signs of perturbation caused by the intrinsic structure of the lens galaxies \citep[][ Hsueh et al., in prep.]{Gilman2016}, more complex mass model models will have to be considered for a robust quantification of dark matter substructure. 

In summary, we have demonstrated in this work that B0712+472 is a similar system to B1555+375 in that the observed flux ratio anomaly can be reproduced by including a disc component in the lens model without the need for additional nearby dark substructures. Thus, baryonic structure can perturb flux ratios of lensed quasars and introduce bias into the inferred substructure abundance.
%when dark substructures contribute no or partial anomalies to the systems. 
Future investigations of the level of intrinsic anomaly from baryonic mass distributions are therefore needed in preparation for the upcoming large scale surveys that are expected to increase  dramatically the sample of lensed AGN.

\section*{Acknowledgments}
JWH and CDF acknowledge the kind hospitality of the MPA during their visits to Garching.  CDF and DJL acknowledge support from NSF-AST-0909119, and CDF further acknowledges support from NSF-AST-1312329.  LVEK is supported in part through an NWO-VICI career grant (project number 639.043.308). LJO acknowledges the STFC for the award of a studentship; MWA also acknowledges support from the STFC in the form of an Ernest Rutherford Fellowship. The NRAO is a facility of the NSF operated under cooperative agreement by Associated Universities, Inc. Based on observations made with the NASA/ESA Hubble Space Telescope, obtained from the data archive at the Space Telescope Science Institute. Some of the data presented herein were obtained at the W. M. Keck Observatory, which is operated as a scientific partnership among the California Institute of Technology, the University of California and the National Aeronautics and Space Administration. The Observatory was made possible by the generous financial support of the W. M. Keck Foundation. The authors wish to recognize and acknowledge the very significant cultural role and reverence that the summit of Mauna Kea has always had within the indigenous Hawaiian community. We are most fortunate to have the opportunity to conduct observations from this mountain.

\bibliographystyle{mnras}
\bibliography{reference}

\bsp
\label{lastpage}

\end{document}